\documentclass{article}
\usepackage{amssymb}
\usepackage{amsmath}
\usepackage{graphicx}
\def\k{\kappa}
\def\sn{\makebox{\,sn}}
\def\cn{\makebox{\,cn}}
\def\dn{\makebox{\,dn}}
\def\sc{\makebox{\,sc}}
\def\dc{\makebox{\,dc}}
\def\nc{\makebox{\,nc}}
\def\s{s}
\def\ss{\hat{s}}
\def\sss{\hat{s}}
\def\tt{\hat{t}}
\def\u{z}
\def\t{t}
\def\r{\vartheta_3^2}
\title{Doubly elliptic strings on the (anti-)de Sitter manifold}
\author{Michel Gaudin$^1$ and Ugo Moschella$^2$ \\
$^1$Institut de physique th\'eorique, CEA, Saclay, 91191 Gif-sur-Yvette - France \\
$^2$Universit\`a degli Studi dell'Insubria - DiSAT, Sezione di Fisica\\
Via Valleggio 11 - 22100 Como - Italy \\
$^2$INFN, Sez di Milano, Via Celoria 16,
20146, Milano - Italy}
\begin{document}
\maketitle
\begin{abstract}{We present a new class of elliptic-like strings on two-dimensional 
manifolds of constant curvature. Our solutions are related to a 
class of identities between Jacobi theta functions 
and to the geometry of the lightcone in one (spacelike) dimension more.}\end{abstract}

\section{Introduction}

The AdS/CFT correspondence has triggered
a revival of interest in the classical Anti-de Sitter (AdS) string equations.
Such equations have been for long known to be classically integrable on general grounds
\cite{Pohlmeyer,devega,Tseytlin:2010jv}.
In the last decade however, several  concrete explicit solutions have been
constructed and mapped into properties of the dual conformal theory \cite{Polyakov,Maldacena}
(for a review and a partial list of references see \cite{Tseytlin:2010jv}).
The various constructions often make use of ingenious procedures,
as for instance the one introduced by Pohlmeyer \cite{Pohlmeyer},
to transform the string equations and the constraints into other kind of equations
for which integration methods already exist, such as the sine-Gordon equation, etc.

In this paper we present a direct construction of certain doubly elliptic
solutions (i.e. solution which are elliptic functions of both the worldsheet coordinates $\t$ and $\s$)
whose intriguing properties seem to have stayed uncovered to date.
These solutions can be obtained in many different ways.
Here we present a construction based on the embedding of the AdS space
into the projective cone in one dimension more.
We exhibit explicit parameterizations for the two-dimensional anti de Sitter, de Sitter and Lobacevskij strings, while we leave the higher dimensional hyperelliptic cases for future investigation \cite{GMP}.

We begin by reformulating the GKP solution \cite{Polyakov}
on the projective cone in one spacelike dimension more.
This mapping put into evidence the deep relation that exists between AdS strings
(and more generally strings on  complex spheres)
and theta functions (and more generally hyperelliptic functions).
A particular role is played by the Virasoro constraints
that are spelled by a class of (perhaps unknown) beautifully simple
identities between theta functions and their derivatives.
Based on these identities we present several interesting new string solutions.
In a related paper we will discuss the general issue
of separating the of variables in the AdS classical strings equations with application to nontrivial examples in higher dimension \cite{GMP}.

\section{Summary of the GKP's rotating string.}
Even though very well known, to motivate what follows we start by briefly reviewing the GKP rigidly rotating folded string.
One considers here a two-dimensional parameterized surface embedded  in $AdS_3$ as follows:
\begin{equation}\left\{\begin{array}{lll}
Y_0 = \cosh \rho(\s) \cos{(\omega_1 \t)}   ,& \ \ Y_1 = \cosh\rho(\s) \sin{(\omega_1 \t)}, \, \cr
Y_2= \sinh \rho(\s) \cos{(\omega_2 \t)}  , &\ \ Y_3 = \sinh\rho(\s) \sin{(\omega_2 \t)},
\end{array}\right. \label{embedding1}
\end{equation}
where
\begin{equation}
AdS_{d}= \{ Y\in {\Bbb R}^{d+1}:\ Y^2 =  Y\cdot Y =
Y_0^2 + Y^2_1- Y^2_2- Y^2_3-\ldots - Y^2_d=1\}. \label{ads}
\end{equation}
The function $\rho$ is required to be periodic:  $\rho(\s) = \rho(\s + 2L)$ where $L$ is an adjustable parameter.

The second step is to derive a  differential equation for $\rho(\s)$
by imposing the conformal gauge constraints:
\begin{equation}
(\partial_\t Y \pm \partial_\s Y)^2  = \omega_1 ^2 \cosh^2 \rho -\omega_2 ^2\sinh^2  \rho -{(\rho')}^2=0. \label{constraint} \end{equation}
Eq. (\ref{constraint}) immediately establishes a relation between the angular velocities $\omega_1$ and $\omega_2$ and the parameter $L$ as follows:
\begin{eqnarray}
2L = \int_0^{2L} d{\s}= 4 \int_0^{\rho_0}\frac{d\rho }{\sqrt{\omega_1 ^2 \cosh^2 \rho -\omega_2 ^2\sinh^2  \rho}} =  4 \int_0^{k}\frac{dx }{\omega_2 \sqrt{1-x^2}\sqrt{k^2-x^2}} = \frac{4 K(k)}{\omega_2}
\end{eqnarray}
where $K(k)$ is the complete elliptic integral of the first kind and  \[\tanh \rho_0 = \pm \frac{\omega_1}{\omega_2}=\pm k\] gives the localization of the two extremal points of the (closed) folded string.

A straightforward computation shows that any  function $\rho(\sigma)$ solving the constraint (\ref{constraint}) automatically
provides a solution of the string equations
\begin{equation}
\partial^2_\t Y_i -  \partial^2_{\s} Y_i + [(\partial_\t Y)^2 -(\partial_{\s} Y)^2] Y_i = 0
\label{string}
\end{equation}
through the embedding (\ref{embedding1}). Two  important conserved quantities are easily  computed in terms of the complete elliptic integrals:
\begin{eqnarray}
{\cal E}& = &\int_0^{2L} (\dot Y_0  Y_1 -  Y_0 \dot Y_1)d\sigma \ =
\frac {4 k E(k)} {1-k^2}   \\
S&=& \int_0^{2L} (\dot Y_2  Y_3 -  Y_2 \dot Y_3)d\sigma =  %\int_0^{2L}  \omega_2\sinh^2 \rho d\s = 4 \int_0^{\rho_0} \frac{ \omega_2\sinh^2 \rho d\rho }{\sqrt{\omega ^2 \cosh^2 \rho -\omega_2 ^2\sinh^2  \rho}}
%= \\ =  4 \int_0^{k} \frac{y^2}{\sqrt{k^2  - y^2 }{(1  - y^2)^{\frac 32} }} dy
 \frac{4E(k)}{1-k^2}  - 4 K(k).
\end{eqnarray}
where $E(k)$ is the complete elliptic integral of the second kind.
%\[ K = \frac{\pi}{2 (1-k^2)^\frac 12 }P_{-\frac{1}{2}}\left(\frac{1+k^2}{1-k^2}\right), \ \ \ \  E = \frac{\pi}{2 (1-k^2)^{\frac 1 4 } }P_{\frac{1}{2}}\left(\frac{2-k^2}{\sqrt{1-k^2}}\right).\]
The above equations provide a relation ${\cal E}={\cal E}(S)$ between  ${\cal E}(k)$ and $S(k)$ (in parametric form) which is relevant to compute the anomalous dimension of twist operators in the dual conformal field theory \cite{Polyakov,pawellek}.

To integrate Eq. (\ref{constraint}) we pose $y=\cosh \rho$.  Eq. (\ref{constraint}) is transformed into the nonlinear differential equation satisfied by a certain Jacobian elliptic function:
\begin{equation}
(y')^2 = \omega_2 ^2\left[-1 + \left(2- {k^2}\right)y^2  -\left(1- {k^2}\right)y^4\right]. \label{constraintg}
\end{equation}
The natural initial condition $\rho(0) = 0$ gives rise to the solution
\begin{equation} \cosh \rho =  \text{nd} \left(\omega \s;k \right), \ \ \ \ \sinh \rho = k \ \text{sd} (\omega \s ;k),
\end{equation}
where we set $\omega_2=\omega$; the string worldsheet is finally parameterized as follows:
\begin{equation}\left\{\begin{array}{lll}
Y_0 = \text{nd} \left(\omega \s ; k \right) \cos{(k \omega \t)}   ,& \ \ Y_1 = \text{nd} \left(\omega \s ;k \right) \sin{(k \omega \t)}, \, \cr
Y_2= k \ \text{sd} (\omega \s;k) \cos{(\omega \t)}  , &\ \ Y_3 = k \ \text{sd} (\omega \s;k) \sin{(\omega \t)}.
\end{array}\right. \label{embeddingfinal}
\end{equation}
A second simple solution is obtained from the latter by the quarter period shift $\omega \s \to  \omega \s +K $:
\begin{equation}
\cosh \rho =  \frac{1}{k'} \ \text{dn} \left(\omega \s;k \right), \ \ \ \ \sinh \rho = \frac{k}{k'} \ \text{cn} (\omega \s;k). \label{gkp2}
\end{equation}
where $k'=\sqrt{1-k^2}$ is the complementary modulus. This completes our review of the GKP solution.
\subsection{Rotating strings on $AdS_5$}
An immediate generalization of the original GKP rigidly rotating string may be obtained
by the following ansatz:
\begin{equation}\left\{\begin{array}{lll}
Y_0 =  b \,\, \dn (\k\s,k) \cos{(\omega t)}, &  Y_1 = b \,\, \dn (\k\s,k) \sin{(\omega t)}, \cr
Y_2 = b_1  \sn (\k\s,k) \cos{(\omega_1 t)},  &  Y_3 = b_1  \sn (\k\s,k) \sin{(\omega_1 t)}, \cr
Y_4 = b_2  \cn (\k\s,k) \cos{(\omega_2 t)},  &  Y_5 = b_2  \cn (\k\s,k) \sin{(\omega_2 t)}.
\end{array}\right. \label{embedding2}
\end{equation}
Two conditions on the three coefficients $b$ follow by imposing $Y\cdot Y=1$ (i.e. $Y\in AdS_5$):
%\begin{eqnarray} \left(b_1^2-b_2^2-k^2 b_3^2\right) \sn^2\k\s +b_2^2+b_3^2=1 \end{eqnarray}
\begin{eqnarray}
 b_1^2 =  {k'}^2 b^2-1\,  , \ \ \ \ \ b_2^2=b^2-1 \,  . \label{condisp}
\end{eqnarray}
Next, we impose the Virasoro constraints $(\partial_\t Y \pm \partial_\s Y )^2   = 0$ and get two more conditions on the angular velocities:
\begin{eqnarray}
 \omega_1^2=\frac{{k'}^2 \left(\left(b^2-1\right) \kappa ^2-b^2 \omega ^2\right)}{1-{k'}^2 b^2}\, , \ \ \ \  \
 \omega_2^2=\frac{\left(1-{k'}^2 b^2\right) \kappa ^2}{b^2-1}+\frac{b^2 \omega ^2}{b^2-1}\, . \label{pp}
\end{eqnarray}
Finally,  the string equations (\ref{string}) give one more condition (at variance with the original GKP model
where the constraints are enough for the string equation to hold, because of its lower dimensionality):
\begin{equation}
%\omega^2 =\left(2 {k'}^2 b^2+k^2-2\right) \kappa ^2, \ \ \
b^2=\frac1 {2 {k'}^2}\left({\frac{\omega^2}{\kappa ^2}+{k'}^2+1 } \right)
\end{equation}
which together with  Eqs. (\ref{condisp}) and (\ref{pp}) implies that

\begin{eqnarray}
& \omega_1^2= %\left(2 {k'}^2 b^2+ {k}^2-1\right)  \kappa ^2 =
\omega^2 + \kappa^2, & \omega_2^2=%\left(2 {k'}^2 b^2-1\right) \kappa ^2 =
\omega^2 + {k'}^2\kappa^2,\\
&  b_1^2 =  \frac 1 {2 }\left({\frac{\omega^2}{\kappa ^2}-{k}^2}\right), & b_2^2=\frac1 {2 {k'}^2}\left({\frac{\omega^2}{\kappa ^2}+{k}^2} \right) . \label{condisp2}
\end{eqnarray}
It is immediately seen that the choice $\omega^2={k}^2\kappa^2$ reproduces the standard GKP solution as given in Eq. (\ref{gkp2}). On the other hand, for any value of the elliptic modulus $k$ there is a one parameter family of rotating GKP-like strings parameterized by an angular velocity $\omega \geq {k}\kappa$.
Now we have three non-zero Cartan conserved quantities, the energy and two spins (we set
$\kappa=1$):
\begin{eqnarray}
{\cal E}_2& = &%\int_0^{2L} (\dot Y_0  Y_5 -  Y_0 \dot Y_5)d\sigma =4 b^2 \int_0^{K} { {\omega}\dn^2 (\s,k) d\sigma } =
\frac {2\omega} { {k'}^2}\left({{\omega^2}+{k'}^2+1 } \right)  E(k^2),
\\
S_1&=& %\int_0^{2L} (\dot Y_1  Y_2 -  Y_1 \dot Y_2)d\sigma =  %4 ({k'}^2 b^2-1)\int_0^{K} { {\omega_1}\sn^2 (\s,k) d\sigma } =
 \frac {2 }{k^2 }\left({{\omega^2}-{k}^2}\right)  \sqrt{\left(\omega^2 + 1\right) }
   \left(K\left({k^2}\right)- E\left({k^2}\right)\right),
\\
S_2&=& %\int_0^{2L} (\dot Y_3  Y_4 -  Y_4 \dot Y_3)d\sigma =%  4 (b^2-1)\int_0^{K} { {\omega_2}\cn^2 (\s,k) d\sigma }=\\ &=&
\frac 2 {{k^2 }{k'}^2}\left({{\omega^2}+{k}^2} \right)\sqrt{\omega^2+{k'}^2  }
   \left(E\left({k^2}\right)-  {k'}^2K\left({k^2}\right)\right),
\end{eqnarray}
which are related by a law ${\cal E}_2 = {\cal E}_2(S_1,S_2)$ that in principle can be obtained by eliminating $\omega$ and $k$;
for $S_1=0$ the function ${\cal E}_2(S_1,S_2)$  reproduces the energy ${\cal E}(S)$  of the standard
GKP string:
\begin{equation}{\cal E} = {\cal E}(S)= {\cal E}_2(0,S)
\end{equation}

\section{The GKP string in homogeneous coordinates}
There is a nice geometrical reinterpretation of the above construction that can be  uncovered by recasting the parametrization (\ref{embeddingfinal}) of the GKP string in terms of theta functions (we adopt the notations of  Whittaker and Watson's classical book \cite{whittaker}).
For the reader's convenience we recall  the basic formulae expressing the Jacobi elliptic functions  as ratios of theta functions and theta constants:
\begin{equation}
\sn(\u,k) =  \frac{\vartheta_3 \vartheta_1(\u/\r)}  {\vartheta_2 \vartheta_4(\u/\r)}, \ \ \
\cn(\u,k) =  \frac{\vartheta_4 \vartheta_2(\u/\r)}  {\vartheta_2 \vartheta_4(\u/\r)}, \ \ \
\dn(\u,k) =  \frac{\vartheta_4 \vartheta_3(\u/\r)}  {\vartheta_3 \vartheta_4(\u/\r)}. \ \ \
\end{equation}
The elliptic moduli $k$ and $k'$ are related to the theta constants as follows:
\begin{equation}
k = \frac {\vartheta_2^2}{\vartheta_3^2}=\frac {\vartheta_2^2(0\,|\,\tau)}{\vartheta_3^2(0\,|\,\tau)} ,  \ \ \ \ k' = \frac {\vartheta_4^2}{\vartheta_3^2}= \frac {\vartheta_4^2(0\,|\,\tau)}{\vartheta_3^2(0\,|\,\tau)}.
\end{equation}
The parametrization (\ref{embeddingfinal}) of the  worldsheet of the GKP string  can then be rewritten as follows:
\begin{equation}\left\{\begin{array}{lll}
Y_0 = \frac{\vartheta_3 \vartheta_4(\sss) }{\vartheta_4 \vartheta_3(\sss)}  \cos{(k \t)},& \ \ Y_1 = \frac{\vartheta_3 \vartheta_4(\sss)}{\vartheta_4 \vartheta_3(\sss)}  \sin{(k  \t)}, \, \cr
Y_2= \frac{\vartheta_2 \vartheta_1(\sss)}{\vartheta_4 \vartheta_3(\sss)} \cos{( \t)}  , &\ \ Y_3 = \frac{\vartheta_2 \vartheta_1(\sss)}{\vartheta_4 \vartheta_3(\sss)}  \sin{( \t)},
\end{array}\right. \label{embeddingfinaluno}
\end{equation}
where we  set $\omega = 1$ and $\hat s=s/\r.$ The above parametrization naturally  suggests the introduction of homogeneous coordinates on the cone
\begin{equation}
C_{2,d}= \{ \xi\in {\Bbb R}^{d+2}:\ \xi^2 =  \xi\cdot \xi =
\xi_0^2 + \xi^2_1- \xi^2_2- \xi^2_3-\ldots - \xi^2_{d+1}=0\} \label{ads}
\end{equation}
(here $d=3$); there we consider the parameterized two-surface
\begin{equation}\left\{\begin{array}{lll}
\xi_0 = {\vartheta_3 \vartheta_4(\sss)}  \cos{(k \t)},& \ \
\xi_1 = {\vartheta_3 \vartheta_4(\sss)} \sin{(k  \t)}, \, \cr
\xi_2= {\vartheta_2 \vartheta_1(\sss)} \cos{( \t)}  , &\ \
\xi_3 ={\vartheta_2 \vartheta_1(\sss)} \sin{( \t)}, \cr
\xi_{4} = {\vartheta_4 \vartheta_3(\sss)};
\end{array}\right. \label{embeddingcone}
\end{equation}
the  string (\ref{embeddingfinal}) on the anti de Sitter GKP is reobtained by taking the ratios $Y_i = \xi_i/\xi_{d+1}$. Note that the condition  $\xi \in C_{2,d}$ just expresses a well-known quadratic identity between theta functions (\cite{whittaker}, p. 466):
\begin{equation}
\xi^2 = \vartheta _3{}^2 \vartheta _4^{}(\sss){}^2-\vartheta _2{}^2 \vartheta _1^{}(\sss){}^2-\vartheta _4{}^2 \vartheta _3^{}(\sss){}^2 = 0. \label{ww}
\end{equation}
The second observation is that the parameterized surface (\ref{embeddingcone}) obeys Virasoro-type constraints on the cone.  The validity of the  constraint $\partial_t \xi \cdot \partial_s \xi = 0$ is immediate. The nontrivial constraint can be shown by using the heat equation which is satisfied by the theta functions (see e.g. \cite{whittaker}). Indeed one has that
\begin{eqnarray}
\partial_s \xi\cdot \partial_s \xi  &=&\frac{\omega^2}{\vartheta _3^4} \left(\vartheta _3{}^2 \vartheta _4^{\prime }(\sss){}^2-\vartheta _2{}^2 \vartheta _1^{\prime }(\sss){}^2-\vartheta _4{}^2 \vartheta _3^{\prime }(\sss){}^2\right) = \cr
&=&\frac{4 {\omega^2}}{\pi i \vartheta _3^4} \left( \frac {\partial \vartheta_3(0|\tau)} {\partial \tau } \vartheta _3{} \vartheta _4(\sss){}^2-\frac {\partial \vartheta_2(0|\tau)} {\partial \tau }  \vartheta _2{} \vartheta _1(\sss){}^2-\frac {\partial \vartheta_4(0|\tau)} {\partial \tau }  \vartheta _4{} \vartheta _3(\sss){}^2\right).
\end{eqnarray}
This equality implies that the function $\partial_s \xi (\cdot) \cdot \partial_s \xi(\cdot)$
is quasi doubly periodic function.
By further observing
that $\partial_s \xi \left(\frac{\pi}2\right) \cdot \partial_s \xi \left(\frac{\pi}2\right)= 0$
one infers that
\begin{eqnarray}
\partial_s \xi\cdot \partial_s \xi = \frac{\omega^2}{\vartheta _3^4} \left(\vartheta _3{}^2 \vartheta _4^{\prime }(\sss){}^2-\vartheta _2{}^2 \vartheta _1^{\prime }(\ss){}^2-\vartheta _4{}^2 \vartheta _3^{\prime }(\sss){}^2\right) = - {\omega^2}\frac{\vartheta _2{}^2 \vartheta _4{}^2} {\vartheta _3{}^2} {\vartheta _2}(\sss){}^2.
\label{thetaidentities}
\end{eqnarray}
On the other hand
\begin{eqnarray}
\partial_t \xi\cdot \partial_t \xi  = {\omega^2}
\frac{\vartheta _2{}^2} {\vartheta _3{}^2} ({\vartheta _2{}^2} \vartheta _4(\sss){}^2-\vartheta _3{}^2 \vartheta _1(\sss){}^2)
={\omega^2} \frac{\vartheta _2{}^2 \vartheta _4{}^2} {\vartheta _3{}^2} {\vartheta _2}(\sss){}^2
\end{eqnarray}
and therefore the constraint
$
\partial_t \xi\cdot \partial_t \xi  +
\partial_s \xi\cdot \partial_s \xi  = 0
$
is  satisfied.

The above result is indeed generally true in the following sense:
suppose that an AdS string is parameterized in terms of inhomogeneous coordinates as follows
\begin{equation}
t,s \to Y_i(t,s)= \frac{\xi_i(t,s)}{\xi_{d+1}(t,s)}, \ \ i=0,1,\ldots, d;
\end{equation}
the map $t,s \to {\xi_\mu(t,s)}$, $\mu=0,1,\ldots, d+1,$ describes a two-surface in $C_{2,d}$.
The condition $\xi^2 = 0$ immediately implies
that
\begin{eqnarray}
\sum _{i=0}^{d-1} {\partial_\u} {Y_i} {\partial_w} {Y_i}
%= \frac {1}{\xi^2_{d+1}} \partial_t{\xi_i} \partial_t {\xi_i}-\frac {2}{\xi^3_{d+1}} (\partial_t {\xi_{d+1}})\,{\xi_i} \partial_t {\xi_i}+\frac {1}{\xi^4_{d+1}} {\xi_i}  {\xi_i} (\partial_t {\xi_{d+1}})^2
= \frac {1}{\xi^2_{d+1}} \sum_{\mu=0}^{d} \partial_\u{\xi_\mu} \partial_w {\xi_\mu}, %\ \ \sum _i {\partial_s} {Y_i} {\partial_s} {Y_i}= \frac {1}{\xi^2_{d+1}} \sum_\mu\partial_s{\xi_\mu} \partial_s {\xi_\mu},\cr \sum _i {\partial_t} {Y_i} {\partial_s} {Y_i}= \sum _\mu \frac {1}{\xi^2_{d+1}} \partial_t{\xi_\mu} \partial_s {\xi_\mu}.
\label{const}
\end{eqnarray}
where $\u,w$ can be either $t$ or $s$. Therefore, if the collection of functions $Y_i$ satisfies the constraints in $AdS_d$, the functions $\xi_\mu$ also do
in $C(2,d)$ and viceversa. This also means that the nontrivial identity (\ref{thetaidentities})
among theta functions and their derivatives is indeed {\em proven}
by exhibiting the map (\ref{embeddingcone}), once the validity
of the Virasoro constraints for the corresponding $AdS$ string is known.
%For example, by rewriting the second GKP solution (\ref{gkp2}) in homogeneous coordinates\begin{equation}\left\{\begin{array}{lll}\xi_0 = {\vartheta_3 \vartheta_3(\sss)}  \cos{(k  \t)},& \ \\xi_1 = {\vartheta_3 \vartheta_3(\sss)} \sin{(k  \t)}, \, \cr\xi_2= {\vartheta_2 \vartheta_2(\sss)} \cos{(  \t)}  , &\\xi_3 ={\vartheta_2 \vartheta_2(\sss)} \sin{( \t)}, \cr\xi_{4} = {\vartheta_4 \vartheta_4(\sss)},\end{array}\right. \label{embeddingcone2}\end{equation}the AdS Virasoro constraints imply a second nontrivial relation among theta functions and their derivatives:\begin{eqnarray}&& \frac{1}{\omega^2}(\partial_t \xi\cdot \partial_t \xi+\partial_s \xi\cdot \partial_s \xi )= \cr &&=\frac{1}{\vartheta _3^4} \left(\vartheta _3{}^2 \vartheta _3^{\prime }(\sss){}^2-\vartheta _2{}^2 \vartheta _2^{\prime }(\sss){}^2-\vartheta _4{}^2 \vartheta _4^{\prime }(\sss){}^2\right) + \frac{\vartheta _2{}^2 } {\vartheta _3{}^2} ({\vartheta _2}{}^2{\vartheta _3}(\sss){}^2-{\vartheta _3}{}^2{\vartheta _2}(\sss){}^2)=0.\label{thetaidentities2}\end{eqnarray}

\section{Doubly elliptic solutions - Preliminaries.}
We are now ready to construct new examples of AdS classical strings.
Here the idea of rotating strings is abandoned in favor of solutions which are  elliptic
functions of both the worldsheet coordinates $t$ and $s$.
These strings have a completely different shape w.r.t.
the rotating examples and may have interesting features
also in view of generalizations to higher dimensions.

Let us describe the typical construction in some detail and consider the following symmetric two-surface embedded in the cone $C_{2,2}$:
\begin{eqnarray} (t,s) \to \left\{\begin{array}{lll}
\xi_0 = {\vartheta_1 (\tt) \, \vartheta_1(\ss)}, & \ \
\xi_1 = {\vartheta_3 (\tt)\, \vartheta_3(\ss)}, \cr
\xi_2= {\vartheta_2 (\tt)\,\vartheta_2(\ss)}  , &\ \
\xi_3 ={\vartheta_4(\tt)\, \vartheta_4(\ss)}.
\end{array}\right. \label{embeddingconeelliptic}
\end{eqnarray}
The vector $\xi$ satisfies the condition $\xi^2=0$ (i.e. $\xi\in C_{2,2}$) because of a well-known quadratic
identity between theta functions of different arguments (\cite{whittaker}, p 487):
\begin{equation}
\xi^2 =  {\vartheta_1(\tt){}^2  \, \vartheta_1(\ss)}{}^2
 - {\vartheta_2 (\tt) {}^2\, \vartheta_2(\ss)}{}^2 + {\vartheta_3 (\tt){}^2 \, \vartheta_3(\ss)}{}^2- {\vartheta_4 (\tt){}^2 \, \vartheta_4(\ss)}{}^2 = 0. \label{funda}
\end{equation}
The symmetry $\xi_\mu (t,s) = \xi_\mu(s,t)$ will reverberate in a certain self-duality of the corresponding strings (we will come back to this point below).
Let us examine the status of the Virasoro constraints:
\begin{equation}
\partial_t \xi \cdot \partial_t \xi + \partial_s \xi \cdot \partial_s \xi=0,\ \ \ \ \ \partial_t \xi \cdot \partial_s \xi=0. \label{vira}
\end{equation}
The validity of the first constraint amounts to establishing another (possibly unknown) identity among theta functions and their derivatives:
\begin{eqnarray}
\partial_t \xi \cdot \partial_t \xi + \partial_s \xi \cdot \partial_s \xi = -\theta_3^{-2}\sum_{i=1}^4 (-1)^\alpha(\vartheta_\alpha'(\tt){}^2  \, \vartheta_\alpha(\ss){}^2
+\vartheta_\alpha(\tt){}^2  \, \vartheta_\alpha'(\ss){}^2)=0. \label{virasoroa}
\end{eqnarray}
This identity may be proven by applying the Laplace operator to  (\ref{funda}):
\begin{eqnarray}
&&0= \frac 12 \, ( \partial_x^2 +\partial^2_y)\sum
_{\alpha=1}^4 (-1)^\alpha(\vartheta_\alpha(x)^{2}\vartheta_\alpha(y){}^2 ) = \cr
&& = \sum_{\alpha=1}^4  (-1)^\alpha(\vartheta'_\alpha(x)^{2}\vartheta_\alpha(y){}^2 + \vartheta_\alpha(x){}^2\vartheta'_\alpha(y)^{2}+ \vartheta_\alpha(x)\vartheta''_\alpha(x)\vartheta_\alpha(y){}^2 + \vartheta_\alpha(x){}^2\vartheta_\alpha(y)\vartheta''_\alpha(y))= \cr && = \sum _{\alpha=1}^4 (-1)^\alpha(\vartheta'_\alpha(x)^{2}\vartheta_\alpha(y){}^2 + \vartheta_\alpha(x){}^2\vartheta'_\alpha(y)^{2})+ \frac{2i}{\pi}\frac{\partial}{\partial \tau}\sum (-1)^\alpha( \vartheta_\alpha(x)^{2}\vartheta_\alpha(y){}^2 ) =
\cr&& =  \sum_{\alpha=1}^4  (-1)^\alpha(\vartheta'_\alpha(x)^{2}\vartheta_\alpha(y){}^2 + \vartheta_\alpha(x){}^2\vartheta'_\alpha(y)^{2} ). \label{virasoro1}
\end{eqnarray}
In the second step we used once more the heat equation.
The other constraint $\partial_t \xi \cdot  \partial_s \xi=0$ is an immediate consequence of the separation of variables  in (\ref{embeddingconeelliptic}).

We can take one step further in understanding Eq. (\ref{virasoroa}) by explicitly computing the quantity
\begin{equation}
\sum_{\alpha=1}^4  (-1)^\alpha\vartheta_\alpha(x)^{2}\vartheta'_\alpha(y){}^2. \label{quantity}
\end{equation}
By Taylor expanding the Jacobi addition formulae (\cite{whittaker}, p 487)
\begin{equation} \vartheta _4{}^2\vartheta _\alpha(y+z) \vartheta _\alpha(y-z) = \vartheta _\alpha(y){}^2\vartheta _4(z){}^2-\vartheta _{{5-\alpha}}(y){}^2\vartheta _1(z){}^2, \ \ \ \alpha = 1,2,3,4. \end{equation}
we rapidly deduce the following differential identities:
\begin{eqnarray}
  \vartheta^{\prime} _\alpha(y){}^2  %\vartheta_\alpha(y) \vartheta''_\alpha(y)-\vartheta _\alpha(y){}^2\frac{\vartheta^{\prime\prime} _4}{ \vartheta _4} +\vartheta _{{5-\alpha}}(y){}^2\left(\frac{\vartheta^{\prime} _1}{\vartheta _4}\right)^2 = \cr
  = \frac{2i}{\pi} \frac{\partial}{\partial \tau}\vartheta_\alpha(y){}^2
  -\vartheta _\alpha(y){}^2\frac{\vartheta^{\prime\prime} _4}{ \vartheta _4}
  +\vartheta _{{5-\alpha}}(y){}^2\left(\frac{\vartheta^{\prime} _1}{\vartheta _4}\right)^2. \label{222ubu}
\end{eqnarray}
By multiplying Eq. (\ref{222ubu}) by $(-1)^\alpha \vartheta_\alpha(x)^{2}$ (where $x$ is an independent variable) and summing over the $\alpha$'s, it follows that
\begin{eqnarray}
  \sum_{\alpha=1}^4 (-1)^\alpha \vartheta_\alpha(x)^{2}\vartheta^{\prime} _\alpha(y){}^2
  = \frac{2i}{\pi}  \sum_{\alpha=1}^4 (-1)^\alpha \vartheta_\alpha(x)^{2} \frac{\partial}{\partial \tau}\vartheta_\alpha(y){}^2
  + \sum_{\alpha=1}^4  (-1)^\alpha \vartheta_\alpha(x)^{2}\vartheta _{{5-\alpha}}(y){}^2\left(\frac{\vartheta^{\prime} _1}{\vartheta _4}\right)^2. \label{222uc}
\end{eqnarray}
On the other hand
\begin{eqnarray}
\frac 12  \partial^2_y  \sum_{\alpha=1}^4  (-1)^\alpha(\vartheta_\alpha(x)^{2}\vartheta_\alpha(y){}^2 ) = 0= \sum_{\alpha=1}^4  (-1)^\alpha \vartheta_\alpha(x){}^2\vartheta'_\alpha(y)^{2}+ \frac{2i}{\pi} \sum_{\alpha=1}^4  (-1)^\alpha\vartheta_\alpha(x)^{2}\frac{\partial}{\partial \tau}( \vartheta_\alpha(y){}^2 )
\end{eqnarray}
Putting everything together we finally get the formula
\begin{eqnarray}
 \sum (-1)^\alpha \vartheta_\alpha(x)^{2}\vartheta^{\prime} _\alpha(y){}^2
  = %\frac 12 \sum (-1)^\alpha \vartheta_\alpha(x)^{2}\vartheta _{{5-\alpha}}(y){}^2\left(\frac{\vartheta^{\prime} _1}{\vartheta _4}\right)^2  =\label{222ud} \cr&& =
  (\vartheta_4(x)^{2}\vartheta _{{1}}(y){}^2-\vartheta_1(x)^{2}\vartheta _{{4}}(y){}^2)\left({\vartheta _2\vartheta _3}\right)^2
  %=  (\vartheta_2(x)^{2}\vartheta _{3}(y){}^2-\vartheta_3(x)^{2}\vartheta _{2}(y){}^2)\left({\vartheta _2\vartheta _3}\right)^2
  \label{22bis}
\end{eqnarray}
where the antisymmetry of the expression (\ref{quantity}) is manifest and  the validity of the constraint equation (\ref{virasoroa}) is confirmed.

\section{Finite open strings.}
\begin{figure}[h]
\includegraphics[width=0.4\textwidth]{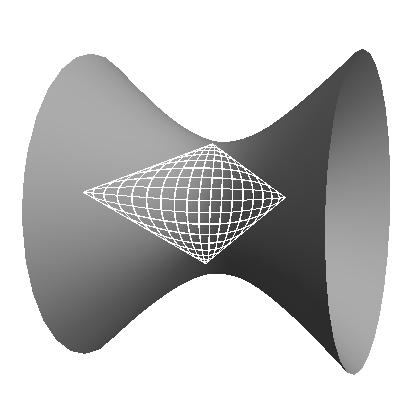}\hfill
\includegraphics[width=0.4\textwidth]{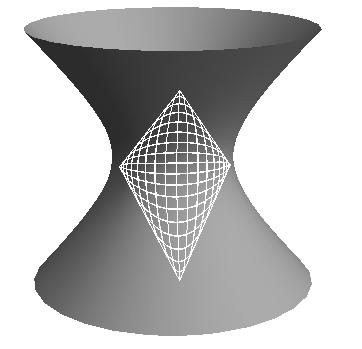}
\caption{A classical string on the  anti de Sitter (resp. de Sitter) manifold, being created at event, reaching its maximal extension,
recontracting and being annihilated at a final event.}
\label{fig1}
\end{figure}

\begin{figure}[h]
\centerline{\includegraphics[width=0.5\textwidth]{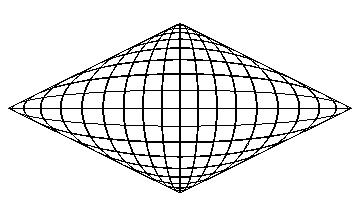}}
\caption{The compact world-surface of an anti de Sitter  classical string (\ref{dsads3})  bordered by four light rays. The diagonals are geodesics.}
\label{fig2}
\end{figure}

Let us now project back the surface (\ref{embeddingconeelliptic}) on the anti de Sitter manifold by singling out
at first the fourth coordinate $\xi_3(\tt,\ss)$.
The map
\begin{equation}
(t,s) \to Y^{(1)}=\left\{\begin{array}{lcl}
Y_0 (t,s) & = & \frac{\xi_0}{\xi_3} =  \frac {\vartheta_1(\tt) \vartheta_1(\ss)}{\vartheta_4(\tt)\vartheta_4(\ss)} = k  \, \sn(t,k)\sn(s,k),  \cr
Y_1 (t,s) & = & \frac{\xi_1}{\xi_3} =  \frac {\vartheta_3(\tt) \vartheta_3(\ss)}{\vartheta_4(\tt)\vartheta_4(\ss)} = \frac{1}{k'} \dn(t,k)\dn(s,k),  \cr
Y_2 (t,s) & = & \frac{\xi_2}{\xi_3} =  \frac {\vartheta_2(\tt) \vartheta_2(\ss)}{\vartheta_4(\tt)\vartheta_4(\ss)}=  \frac k {k'} \cn(t,k)\cn(s,k),  \cr
\end{array}\right. \label{dsads3}
\end{equation}
is an embedding of a two-surface in the two-dimensional anti-de Sitter manifold
$AdS_2$ (see Fig. \ref{fig1}).
Eq. (\ref{dsads3}) represents equally well a world-sheet on the two-dimensional de Sitter universe $dS_2$, the two manifolds coinciding in dimension two. In both cases the Virasoro constraints are satisfied by Eqs. (\ref{const}) and (\ref{virasoroa}).
As for the  GKP string,  the validity of the constraints
is enough for the string equations to hold (this fact is of course not true in general dimension).
We can  elaborate on this as follows:
a concise way of writing the second derivatives of the $Y_i$'s is
\begin{eqnarray}
&& \partial^2_\u  Y_i = ( 2k^2\sn(\u,k)^2-\kappa_i^2 )Y_i \label{bb}
\end{eqnarray}
where $\u$ denotes either  $t$ or  $\s$ and
$\kappa_0^2= 1+k^2$, $\kappa_1^2=1$,  $\kappa_2^2= k^2$. From Eqs. (\ref{const}) and (\ref{22bis}) we get that
\begin{eqnarray}
(\partial_t  Y)^2  =
 \frac{(\vartheta_1(\ss)^{2}\vartheta _{{4}}(\tt){}^2-\vartheta_4(\ss)^{2}\vartheta _{{1}}(\tt){}^2)\left({\vartheta _2/\vartheta _3}\right)^2}{\vartheta_4{}^2(\tt)\vartheta_4{}^2(\ss)} =  k^2 \left(\text{sn}\left(s,k\right)^2-\text{sn}\left(t,k\right)^2\right)=-(\partial_\s  Y)^2 \label{deriv}
\end{eqnarray}
so that
\begin{equation}
\partial^2_{\u}Y_i + (\partial_\u Y)^2 Y_i = \left(k^2\text{sn}\left(s,k\right)^2 +k^2\text{sn}\left(t,k\right)^2 -\kappa_i^2\right)Y_i.
\end{equation}
The symmetry of the rhs in the exchange of $t$ and $s$ finally shows that the string equations (\ref{string}) are satisfied.
The worldsheet of the string is a compact surface  symmetric by reflection w.r.t. the two  arcs of geodesics joining its four vertices.
The worldsheet is bordered by the four light rays joining the same vertices (see Figures \ref{fig1} and \ref{fig2}).
A perhaps surprising feature  of the above symmetric solution is that the family of curves $\s \to Y(\t,\s)$ (labeled by fixed values of $\t$) and the family of curves $\t \to Y(\t,\s)$ (labeled by fixed values of $\s$) coincide. Any such curve is a closed (i.e periodic) curve made by four arcs, two spacelike and two timelike; the four distinguished points where the tangent to the curve becomes lightlike move at the speed of light and  may be interpreted as the string endpoints (see Figure \ref{fig2}). By Eq. (\ref{deriv})
those endpoints are identified by the conditions
\begin{equation}
t = s, \;\; \;2 K(k)  - s,\;\;\; 2 K(k)  + s, \;\;\; 4 K(k)  - s,
\end{equation}
where the elliptic integral $K(k)$ is the quarter period.
A possible physical interpretation of this solution is that of an open string created at
a certain event of the AdS spacetime and expanding up to a maximal length before starting to recontract. The configuration of maximal length (the diagonal)
is the only one that coincides with a spacelike geodesic. At this moment the endpoints  suddenly invert their motion and the string starts recontracting to be finally annihilated at an event which is the mirror image of the event where the string was created w.r.t the plane containing the diagonal spacelike geodesic (and the geometrical center of the AdS manifold). The midpoint of the string is at rest (i.e moves along a timelike geodesic) and we may use it to measure the proper lifetime of the string
\begin{equation}
t(k) = 2 \arcsin k = \arccos (1-2k^2)
\end{equation}
The maximal  length reached by the string is given by
\begin{equation}
l(k) =  {\rm arccosh} \left( \frac{1+k^2}{1-k^2}\right).
\end{equation}
 The action -- the area of the worldsheet -- has an intrinsic geometric meaning:
\begin{eqnarray}
{\cal A} &=& \int \sqrt{h} dt ds =   k^2 \int \sqrt{\left(\text{sn}\left(t,k\right)^2-\text{sn}\left(s,k\right)^2\right)^2} dt ds=%4k^2 \int_0^{2K} \int_{0}^{t} {\left(\text{sn}\left(t,k\right)^2-\text{sn}\left(s,k\right)^2\right)} dt ds
\cr&= &
8 k^2 \int_0^{2K} t \  \text{sn}\left(t,k\right)^2 dt - 8 (K(k)-E(k)) K(k)
= 8 (K(k)-E(k)) K(k) .
\end{eqnarray}
A second possible interpretation of this solution is that of a folded closed string.
Suppose indeed that there are two sheets; each time that the string hits the lightlike boundary passes
to the other sheet and it closes back after a full period.
The other sheet may thought as copies of the same AdS universe.
In this case when passing to the second sheet the curve become timelike
and then again spacelike and once more timelike before closing.
Or either, it also is conceivable that the second sheet is a de Sitter one.
In this case the four arcs are all spacelike and there
are four points where the tangent to curve become lightlike. In both cases
the following (folded) surface may be ascribed to the string:
\begin{eqnarray}
{\cal A}' =    k^2 \int {\left(\text{sn}\left(t,k\right)^2-\text{sn}\left(s,k\right)^2\right)}dt \wedge ds=
16 k^2 \int_0^K \text{sn} \left(t,k\right)^2 dt \int_{0}^{K}  ds = 16 (K(k)-E(k))K(k) \cr
\end{eqnarray}
and obviously ${\cal A}'= 2{\cal A}$.

When
\begin{equation}
k = k_n = \sin({\pi}/{n}), \ \ \ n= 2,3,\ldots
\end{equation}
the above process can be repeated to produce an oscillating string going from zero to its maximal extension and back forever on the anti de Sitter periodic manifold. If we drop the above quantization condition we may always consider a string  oscillating forever but we have to move on the covering of the AdS manifold.

\begin{figure}[h]
{\includegraphics[width=0.5\textwidth]{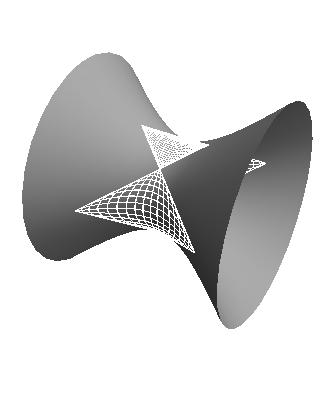}}\hfill
{\includegraphics[width=0.5\textwidth]{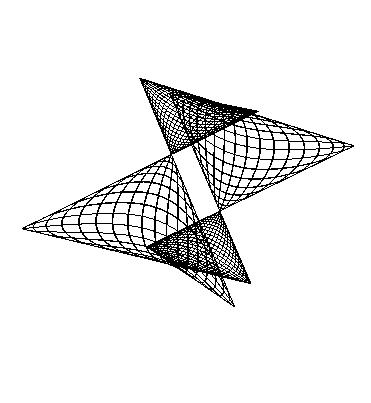}}
%\centerline{\includegraphics[width=0.4\textwidth]{figura4c.eps}}
\caption{An oscillating periodic AdS solution. every time the string reaches a conical point the elliptic modulus jump to the complementary one.
The  boundary of the crown-shaped string worldsheet is made by eight lightlike vectors belonging to AdS bulk.}
\label{fig3}
\end{figure}

The right way is however to continue the solution (\ref{dsads3}) with a second one obtained by the substitution $k\to k'$. The so-obtained  string is a time periodic solution on the true AdS manifold  (as opposed to its covering - see Figure \ref{fig3}). The total action corresponding to this string is now given by \begin{eqnarray}
{\cal A} &=& \int \sqrt{h} dt ds =    8 [K(k)^2+K(k')^2-E(k) K(k) - E(k') K(k')].
\end{eqnarray}

\section{Semi-infinite strings}
Let us now  single out the third coordinate $\xi_2(\tt,\ss)$;
we get another parameterized two-surface on $AdS_2$ (or $dS_2$), always satisfying the Virasoro constraints:
\begin{equation}
(t,s) \to Y^{(2)}(t,s;k) =  \left\{\begin{array}{lcl}
Y_0 (t,s) & = & \frac{\xi_0}{\xi_2} =  \frac {\vartheta_1(\tt) \vartheta_1(\ss)}{\vartheta_2(\tt)\vartheta_2(\ss)} =  {k'}  \, \sc(t,k)\sc(s,k),  \cr
Y_1 (t,s) & = & \frac{\xi_1}{\xi_2} =  \frac {\vartheta_3(\tt) \vartheta_3(\ss)}{\vartheta_2(\tt)\vartheta_2(\ss)} = \frac 1 {k} \dc(t,k)\dc(s,k),  \cr
Y_2 (t,s) & = & \frac{\xi_3}{\xi_2} =  \frac {\vartheta_4(\tt) \vartheta_4(\ss)}{\vartheta_2(\tt)\vartheta_2(\ss)}=  \frac {k'} {k} \nc(t,k)\nc(s,k);  \cr
\end{array}\right. \label{dsads2}
\end{equation}
The second derivatives of the $Y_i$'s have again a simple structure:
\begin{eqnarray}
&& \partial^2_\u  Y_i = ( 2\, \text{dc}(\u,k)^2-\kappa_i^2 )Y_i \label{bb}
\end{eqnarray}
where $\u$ denotes either $\s$ or $t$ and
$\kappa_0^2=k^2$, $\kappa_1^2=1+k^2$,  $\kappa_2^2= 1$.
From Eqs. (\ref{const}) and (\ref{22bis})
\begin{eqnarray}
(\partial_t  Y)^2  = -(\partial_\s  Y)^2 =
 %\frac{(\vartheta_3(\ss)^{2}\vartheta _{2}(\tt){}^2-\vartheta_2(\ss)^{2}\vartheta _{3}(\tt){}^2)\left({\vartheta _2/\vartheta _3}\right)^2}{\vartheta_2{}^2(\tt)\vartheta_2{}^2(\ss)} =  \\=
  \text{dc}\left(s,k\right)^2-\text{dc}\left(t,k\right)^2.\label{deriv2}
\end{eqnarray}
and therefore
\begin{equation}
\partial^2_{\u}Y_i + (\partial_\u Y)^2 Y_i = \left(\text{dc}\left(s,k\right)^2 +\text{dc}\left(t,k\right)^2 -\kappa_i^2\right)Y_i.
\end{equation}
Once again the symmetry of the rhs in the exchange of $t$ and $s$ shows that that (\ref{dsads2}) solves the string equations (\ref{string})
on the two-dimensional anti-de Sitter spacetime, or well on the two-dimensional de Sitter spacetime.
The family of curves $\s \to Y(\t,\s)$ (labeled by fixed values of $\t$) and the family of curves $\t \to Y(\t,\s)$ (labeled by fixed values of $\s$) coincide. Any such curve is an infinite curve made by three arcs: two spacelike and one timelike in the AdS case; two timelike and one spacelike in the dS case; the two distinguished points where the tangent to the curve becomes lightlike move at the speed of light and  may be interpreted as the string endpoints.

\begin{figure}[h]
{\includegraphics[width=0.4\textwidth]{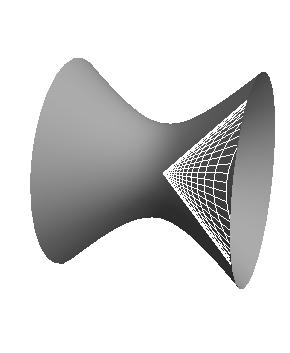}}\hfill
{\includegraphics[width=0.3\textwidth]{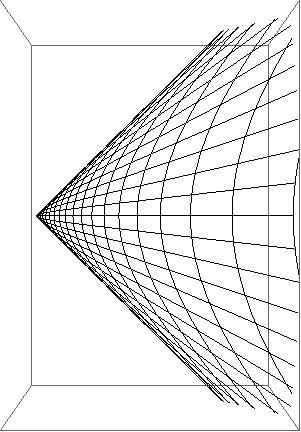}}
\caption{The anti de Sitterian semi-infinite string (\ref{dsads2}) extending towards the spacelike boundary. }
\label{fig4}
\end{figure}

A possible physical interpretation of the  string is as follows: in the anti de Sitter case  we may think of a semi-infinite open string. One endpoint is at spacelike infinity, the other moves at the speed of light towards an event A. When the string configuration coincides with a spacelike geodesics, the second endpoint suddenly inverts the direction of its speed and the string goes away to infinity (see Figure \ref{fig4}). The de Sitter interpretation of this solution is that of a finite open string being created at an event A and expanding forever to timelike infinity (or well a string coming in from minus infinity to be annihilated at the event A).

\begin{figure}[h]
{\includegraphics[width=0.4\textwidth]{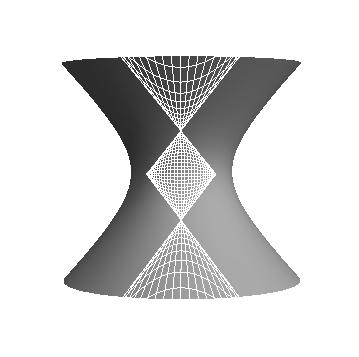}}
\hfill{\includegraphics[width=0.4\textwidth]{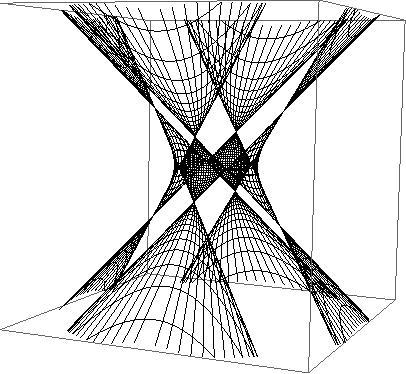}}
\caption{We represent here the complete solution in the de Sitter case.
At the left  a string contracting to a point, expanding and recontracting again, and then expanding forever. At the right the complete (real) solution. Every time that the string passes through a singular point the elliptic modulus jumps to the complementary modulus: the four patches contiguous to a given patch of elliptic modulus $k$ all have modulus $k'$.}
\label{fig5}
\end{figure}

The strings (\ref{dsads3}) and (\ref{dsads2}) can be merged into a single string
made by three disconnected parts that touch at the precise moment where the endpoints
invert their speed. As before the natural choice is to glue together the solution (\ref{dsads3})
at the elliptic modulus $k$ with the solution (\ref{dsads2}) at $k'$.
Indeed such strings are the real manifolds of a unique complex string that lives on the complex two sphere:
\begin{equation}
Y^{(2)}(t,s;k') = Y^{(1)}(it,is;k)
\end{equation}
Putting everything together we finally obtain the string represented in Figure \ref{fig5}.

\section{Euclidean strings.}
The above strings exhaust the possible open strings on the two dimensional anti de Sitter (resp. de Sitter) manifold.
What about the two remaining possible quotients? They will not give rise to any new solution. Indeed let us single out the second coordinate $\xi_1(\tt,\ss)$; we get
\begin{equation}
(t,s) \to \left\{\begin{array}{lclll}
Y_0 (t,s) & = & \frac{\xi_2}{\xi_1} = k  \, \text{cd}(t,k)\text{cd}(s,k) &=&  k  \, \text{sn}(t+K,k)\text{sn}(s+K,k),  \cr
Y_1 (t,s) & = & \frac{\xi_3}{\xi_1}  ={k'} \ \text{nd}(t,k)\text{nd}(s,k)&=& \frac 1{k'} \, \text{dn}(t+K,k)\text{dn}(s+K,k),    \cr
Y_2 (t,s) & = & \frac{\xi_0}{\xi_1} =    {k'} {k}\  \text{sd}(t,k) \text{sd}(s,k)&=& \frac k {k'}\, \cn(t+K,k)\cn(s+K,k),  \cr
\end{array}\right. \label{dsads}
\end{equation}
i. e. this solution is obtained from (\ref{dsads3}) by a quarter period shift.
\begin{equation}
Y^{(3)}(t,s;k') = Y^{(1)}(t+K,s+K;k)
\end{equation}
Similarly
%\begin{equation}(t,s) \to \left\{\begin{array}{lcl}Y_0 (t,s) & = & \frac{\xi_2}{\xi_0} =  \frac{1}{k'} \text{cs}(t,k)\text{cs}(s,k)={k'}  \, \sc(t+K,k)\sc(s+K,k),      \crY_1 (t,s) & = & \frac{\xi_3}{\xi_0}  = \frac 1 k  \, \text{ns}(t,k)\text{ns}(s,k)=\frac 1 {k} \dc(t+K,k)\dc(s+K,k),  \crY_2 (t,s) & = & \frac{\xi_1}{\xi_0} =  \frac 1 {k k'} \text{ds}(t,k)\text{ds}(s,k) =  \frac {k'} {k} \nc(t+K,k)\nc(s+K,k); \cr\end{array}\right. \label{dsads}\end{equation}
the solution obtained by singling out the first coordinate $\xi^0$ coincides with (\ref{dsads2}) by a quarter period shift:
\begin{equation}
Y^{(4)}(t,s;k') = Y^{(2)}(t+K,s+K;k).
\end{equation}
As a final remark we notice that by replacing $t$ with $i t$ in Eq. (\ref{dsads}) we get that the embedding
\begin{equation}
(t, s) \to \left\{\begin{array}{lcl}
X_0 (t,s)& = & \frac 1 {\sqrt{1-k^2}} \makebox{dc}\left(t,k'\right) {\makebox{dn}\left(s,k\right) } =  \frac {\vartheta_2}{\vartheta_4}
\frac{\vartheta_3(\tt)}  {\vartheta_2(\tt)} \frac{ \vartheta_3(\ss)}  { \vartheta_4(\ss)}.\cr
X_1 (t,s) & = & \frac{k}{k'} \makebox{nc}\left(t,k'\right) \makebox{cn}\left(s,k\right)  =  \frac {\vartheta_2}{\vartheta_4}
 \frac {\vartheta_4(\tt)} {\vartheta_2(\tt)}\frac{\vartheta_2(\ss)}  { \vartheta_4(\ss)}, , \ \ \
\cr
X_2 (t,s) &=& k \, \makebox{sc} \left(t,k'\right) \makebox{sn}\left(s,k\right) = \frac {\vartheta_2}{\vartheta_4} \frac{\vartheta_1(\tt)}  { \vartheta_2(\tt)} \frac{ \vartheta_1(\ss)}  { \vartheta_4(\ss)},
\end{array}\right.
\end{equation}
represents either a closed string expanding or an open string oscillating in a Lobacevsky space, identified here with the Euclidean AdS manifolds. Identical result is obtained by the replacement
$s$ with $i s$. Thus our meethod provides also solution for the hyperbolic sigma-model that are real. On the other hand, as it is well known, this is not possible for the corresponding sigma-model on the sphere, subject to the constraints (\ref{vira}) \cite{mira}.
%\end{document}
\section{Open/closed self-dual strings}
\begin{figure}[h]
{\includegraphics[width=0.4\textwidth]{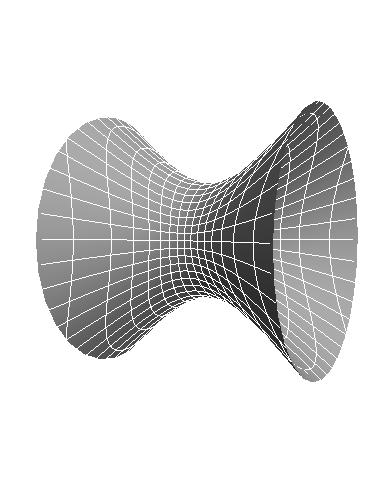}} \hfill
{\includegraphics[width=0.5\textwidth]{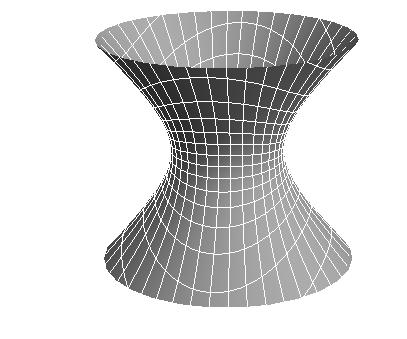}}
\caption{At the left: the infinite AdS solution. At the right: the solution wrapping around the de Sitter space.
In both cases no point moves at the speed of light and the string extends all over the whole spacetime.}\label{fig7}
\end{figure}

There is a second well-known relations between theta functions (\cite{whittaker}, p 487) that may be mapped into solutions of the string equations on $AdS_2$ (or either $dS_2$):
\begin{equation}
\xi^2 =  {\vartheta_1(\tt){}^2  \, \vartheta_3(\ss)}{}^2 + {\vartheta_2 (\tt){}^2 \, \vartheta_4(\ss)}{}^2
 - {\vartheta_3 (\tt) {}^2\, \vartheta_1(\ss)}{}^2 - {\vartheta_4 (\tt){}^2 \, \vartheta_2(\ss)}{}^2 = 0. \label{fund}
\end{equation}
The associated two-surface embedded in the cone $C_{2,2}$ is now given by:
\begin{eqnarray} t,s \to \left\{\begin{array}{lll}
\xi_0 = {\vartheta_1 (\tt) \, \vartheta_3(\ss)}, & \ \
\xi_1 = {\vartheta_2 (\tt)\, \vartheta_4(\ss)}, \cr
\xi_2= {\vartheta_3 (\tt)\,\vartheta_1(\ss)}  , &\ \
\xi_3 ={\vartheta_4(\tt)\, \vartheta_2(\ss)},
\end{array}\right. \label{embeddingconeelliptic2bis}
\end{eqnarray}
and the embedding
\begin{equation}
(\t, \s) \to \left\{\begin{array}{lclcl}
Y_0 (\t,\s)& = & \frac{\xi_0}{\xi_3} &=& {\text{sn} (\t,k)}\ { \text{dc} (s,k)}
, \cr
Y_1 (\t,\s) & = & \frac{\xi_1}{\xi_3} &=& {\text{cn} (\t,k)}\ { \text{nc} (s,k)}
, \cr
Y_2 (\t,\s)& = & \frac{\xi_2}{\xi_3} &=& {\text{dn} (\t,k)}\ { \text{sc} (s,k)}
\end{array}\right. \label{dsads2bis}
\end{equation}
represents a string. The anti-de Sitter interpretation is that of an infinite string vibrating periodically. No point moves at the speed of light and therefore the string extends to infinity.
Exchanging the roles of $\t$ and $\s$ the above solution may also be sees as a closed string wrapping around a two-dimensional de Sitter spacetime and undergoing a (cosmological) process contraction followed by an expansion, while vibrating. The equal time section of the string are however not cosmological sections of the de Sitter manifold (see figure \ref{fig7}).
The other possible ratios of the homogeneous coordinates  do not produce any new solution as it may be understood by the absence of a boundary of the worldsheet in either the  AdS or the dS case or circling around the manifold. For instance
\begin{equation}
(\t, \s) \to \left\{\begin{array}{lclcl}
Y_0 (\t,\s)& = & \frac{\xi_0}{\xi_2} &=& {\text{sd} (\t,k)}\ { \text{ds} (s,k)} = {\text{cn} (\t+K,k)}\ { \text{nc} (s+K,k)}
, \cr
Y_1 (\t,\s) & = & \frac{\xi_1}{\xi_2} &=& {\text{cd} (\t,k)}\ { \text{ns} (s,k)}
={-\text{sn} (\t+K,k)}\ { \text{dc} (s+K,k)} \cr
Y_2 (\t,\s)& = & \frac{\xi_3}{\xi_2} &=& {\text{nd} (\t,k)}\ { \text{cs} (s,k)} = - {\text{dn} (\t+K,k)}\ { \text{sc} (s+K,k)}
\end{array}\right. \label{dsads2bis}
\end{equation}
and the solution is obtained from the previous simply by a quarter period shift in both the worldsheet coordinates.

\section{Conclusions and Outlook}
We have presented a class of doubly elliptic solution of the string equations  on the de Sitter and the Anti-de Sitter manifold
The main ingredients of our treatment are the separation of variables in the string equations and the relation of the model with the algebraic structure of the Jacobian elliptic theta functions. In the present context that relation  is made clear through the embedding of the strings in the projective cone.
The constant curvature manifold arise as quotient of that cone.
In a forthcoming work \cite{GMP} we will hopefully present a general treatment of the separation of variables
not limited to the two-dimensional case as well as nontrivial solutions of the string equation.
\section*{Acknowlegdments} Several enlightening discussions with Vincent Pasquier are gratefully acknowledged.
U.M. Thanks the Perimeter Institute for Theoretical Physics, the Institut de Physique Theorique, CEA-Saclay and the IHES  for warm hospitality and support.


\begin{thebibliography}{99}
%\cite{Pohlmeyer:1975nb}
\bibitem{Pohlmeyer}
  K.~Pohlmeyer,
  ``Integrable Hamiltonian Systems And Interactions Through Quadratic
  Constraints,''
  Commun.\ Math.\ Phys.\  {\bf 46} (1976) 207.
  %%CITATION = CMPHA,46,207;%%
\bibitem{devega} H. J. De Vega and N. Sanchez, ``Exact integrability of strings in $d$-dimensional de Sitter
space-time'',Phys. Rev. D47, 3394 (1993).
%\cite{Tseytlin:2010jv}
\bibitem{Tseytlin:2010jv}
  A.~A.~Tseytlin,
  ``Review of AdS/CFT Integrability, Chapter II.1: Classical AdS5xS5 string solutions,''
  Lett.\ Math.\ Phys.\  {\bf 99}, 103 (2012)
  [arXiv:1012.3986 [hep-th]].
  %%CITATION = ARXIV:1012.3986;%%
%\cite{Gubser:2002tv}
\bibitem{Polyakov}
  S.~S.~Gubser, I.~R.~Klebanov and A.~M.~Polyakov,
  ``A Semiclassical limit of the gauge / string correspondence,''
  Nucl.\ Phys.\ B {\bf 636}, 99 (2002)
  [hep-th/0204051].
  %%CITATION = HEP-TH/0204051;%%
  %\cite{Alday:2007hr}
\bibitem{Maldacena}
  L.~F.~Alday and J.~M.~Maldacena,
  ``Gluon scattering amplitudes at strong coupling,''
  JHEP {\bf 0706}, 064 (2007)
  [arXiv:0705.0303 [hep-th]].
  %%CITATION = JHEPA,0706,064;%%
%\cite{Pawellek:2011xd}
\bibitem{GMP}
M. Gaudin, U. Moschella, V. Pasquier, in preparation.
\bibitem{pawellek}
  M.~Pawellek,
  %``Semiclassical Strings in AdS$_5 \times S^5$ and Automorphic Functions,''
  Phys.\ Rev.\ Lett.\  {\bf 106}, 241601 (2011)
  [arXiv:1103.2819 [hep-th]].
  %%CITATION = ARXIV:1103.2819;%%\end{document}
\bibitem{whittaker}
E. T. Whittaker and G. N. Watson, {\em A course in modern analysis.} Fourth Edition, Cambridge University Press, Cambridge, 1927.
%\cite{Miramontes:2008wt}
\bibitem{mira}
  J.~L.~Miramontes,
  %``Pohlmeyer reduction revisited,''
  JHEP {\bf 0810}, 087 (2008)
  [arXiv:0808.3365 [hep-th]].
  %%CITATION = ARXIV:0808.3365;%%


\end{thebibliography}
\end{document}